\documentclass[structabstract]{aa}  
\usepackage{graphicx}
\usepackage{txfonts}
\input{epsf}
\begin{document}
   \title
{The Evolution of Early-type Galaxies Selected by Their Spatial Clustering}
\titlerunning{Evolution of Early-type Galaxies Selected by Clustering}


   \author{Nelson Padilla
          \inst{1} \and
	  Daniel Christlein\inst{2}\and Eric Gawiser\inst{3}
          \and
          Danilo Marchesini\inst{4}
          }

   \institute{
 Departamento de Astronom\'\i a y Astrof\'\i sica, and Centro de Astro-Ingenier\'\i a, Pontificia
     Universidad Cat\'olica de Chile, V. Mackenna 4860, Santiago 22, Chile.\\
              \email{npadilla@astro.puc.cl}
         \and
 Max Planck Institut fur Astrophysik, Karl-Schwarzschild-Str. 1
Postfach 1317 D-85741 Garching, Germany\\
         \and
 Department of Physics and Astronomy,
     Rutgers University, 136 Frelinghuysen Rd, Piscataway, NJ 08854, USA.\\
         \and
 Physics \& Astronomy Department,
     Tufts University, Robinson Hall, Room 257, Medford, MA 02155, USA.\\
             }

   \date{Received September 15, 1996; accepted March 16, 1997}

 
  \abstract
   {
   }
   {
We present a new method that uses luminosity or stellar mass functions combined with
clustering measurements to select samples of galaxies at different
redshifts likely to follow a progenitor-to-descendant relationship.  As the method
uses clustering information, we refer to galaxy samples selected this way as clustering-selected samples.
We apply this method to infer the number of mergers during
the evolution of MUSYC early-type galaxies (ETGs) 
from $z\simeq 1$ to the present-day.
   }
   {
The method consists in using clustering information
to infer the typical dark-matter halo mass of the hosts of the selected progenitor galaxies.
Using $\Lambda$CDM predictions, it is then possible to follow these haloes to a later time
where the sample of descendants will be that with the clustering of these descendant haloes.
   }
   {
This technique shows that ETGs at a given redshift evolve into 
brighter galaxies at lower redshifts (considering rest-frame, passively evolved optical 
luminosities).  This
indicates that the stellar mass of these galaxies increases with time and that, in principle,
a stellar mass selection at different redshifts does not provide samples of galaxies
in a progenitor-descendant relationship.  
   }
   {
The
comparison between high redshift ETGs and their likely
descendants at $z=0$ points to
a higher number density for the progenitors  by a factor $5.5\pm4.0$,
implying the need for mergers to decrease their number density by today.
Because the luminosity densities of progenitors and descendants
are consistent, our results show no need for significant
star-formation in ETGs since $z=1$, which indicates that the needed mergers are dry,
i.e. gas free.
   }


   \maketitle
%

\section{Introduction}
\label{sec:intro}

The study of the population of early type galaxies (ETGs)
at different redshifts has been used extensively to test our
knowledge of the galaxy formation process, in particular, the assembly
of the stellar content of massive galaxies.  One of the main advantages
in studying ETGs resides in their simple evolution, which allows to use passive
evolution to infer their likely $z=0$ luminosities taking into account the aging
of their stellar populations; this allows to use passively evolved luminosities (PEL) 
as good proxies for their stellar mass.
Analyses of the evolution of the stellar mass and PEL functions
have found that high stellar mass ($M_s>10^{11}$h$^{-1}M_{\odot}$),
passive galaxies do not show evolution
in their comoving space density since $z\sim 1$
(Cimatti et al., 2002, 2004; McCarthy et al., 2004;
Glazebrook et al. 2004; Daddi et al., 2005; Saracco et al., 2005;
Bundy et al., 2006;
P\'erez-Gonz\'alez et al., 2008; 
Marchesini et al., 2009; { Ferreras et al., 2009}).
This result has been interpreted as evidence that
the stellar content of such galaxies is already in place at high redshifts,
ruling out the involvement of mergers (even dry i.e., gas-free) since $z\sim 1$.

The details of the evolution of ETGs are still in debate.
Galaxy formation models embedded in hierarchical cosmologies, such as the
$\Lambda$-Cold Dark Matter ($\Lambda$CDM) model, have been compared to observational
data from different redshifts and wavelength ranges, via numerous statistical probes; 
in some cases the agreement
appears to be excellent and in other cases there seem to be important discrepancies.
Some earlier failures of the models
have been due to the consequences of adopting the wrong cosmological model
(see for instance Heyl et al., 1995), and others
to the lack of important astrophysical processes in the galaxy formation recipe, such
as for instance, galaxy mergers (e.g. Cole et al., 1994) and
feedback from AGN which when included solves discrepancies in the number density
of bright galaxies (e.g. Bower et al., 2006).  
More recent models show reasonable agreement with galaxies in the local Universe 
(Kauffmann et al., 1999; Cole et al., 2000; Bower et al., 2006; De Lucia et al., 2006; Lagos, Cora \& Padilla, 2008;
{ Khochfar \& Ostriker, 2008}; Lagos, Padilla \& Cora, 2009), but still struggle to agree with a number of 
recent statistical measurements of $z\sim1$ to $3$ galaxies,
such as the stellar mass function (Perez-Gonzalez et al., 2008; Marchesini et al., 2009;
but see Benson \& Devereux, 2010),
the galaxy luminosity function (Marchesini \& van Dokkum, 2007), 
and SCUBA (Holland et al., 1999) number counts of
$z\simeq 3$ galaxies detected in sub-millimetre bands, so far only succesfully reproduced
by the Baugh et al. (2005) model.  A slightly better agreement is found for $z\sim 1$
galaxies as in the comparison between model and observed Luminous Red Galaxies by Almeida
et al. (2008).

Hopkins et al. (2008) presented an analysis of the evolution of the density of ETGs
from a theoretical point of view.  They show that part of the apparent mismatch between
models and observations could originate from an inadequate mechanism to quench
the star-formation in galaxies which have experienced major mergers.  Due to this
neither the Croton et al. (2006) nor the Bower et al. (2007) star-formation quenching approaches,
via a dark-matter host-halo mass threshold or disc instabilities, respectively,
are able to reproduce the $z\sim1-3$ measurements of the number density of bright galaxies.  
Hopkins et al. suggest an intermediate solution may fit observational data.

It could also be possible that
the disagreement between models and observations 
is produced by systematic biases from large redshift measurement errors or selection effects affecting the
observational statistics at high redshifts, or simply due to the different and unknown nature of
high redshift galaxies.  
Some results indicate that 
the effect from large redshift errors does not seem to be a major factor. For instance 
the analysis of high redshift clusters in the infrared
by Mancone et al. (2010) shows an evolution of $1.5$ magnitudes between
redshifts of $1$ and $0$, which is consistent with passive evolution,
even when clusters allow a better handle on redshift estimate errors.
This conclusion was also obtained by Muzzin et al. (2008) for the GCLASS cluster survey at z=1.
On the other hand, 
Marchesini et al. (2010) showed that allowing for the existence of a previously unrecognized 
population of massive, old, and very dusty galaxies at $z\sim2.6$ would result in good agreement 
between the observed abundance of massive galaxies at $z=3.5$ and that predicted by semi-analytic 
models.

\begin{figure*}
\begin{picture}(450,365)
\put(-30,140){\includegraphics[width=20.cm,clip]{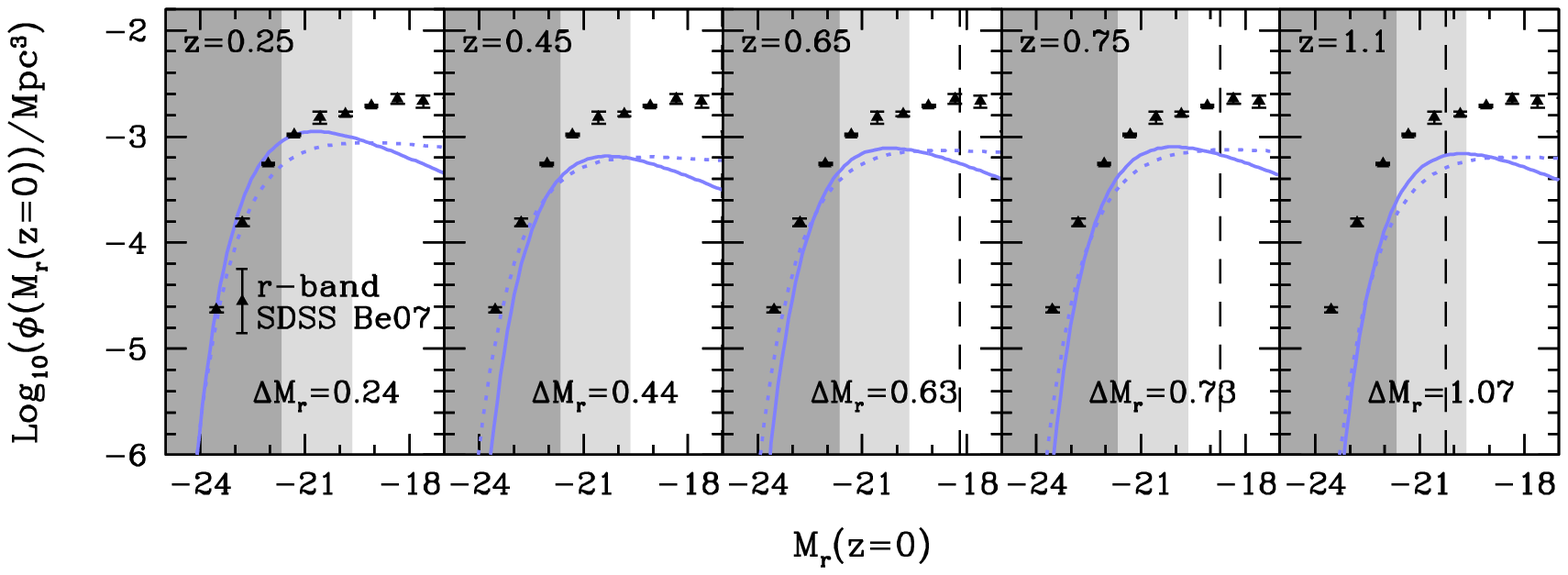}}
\put(-30,-50){\includegraphics[width=20.cm,clip]{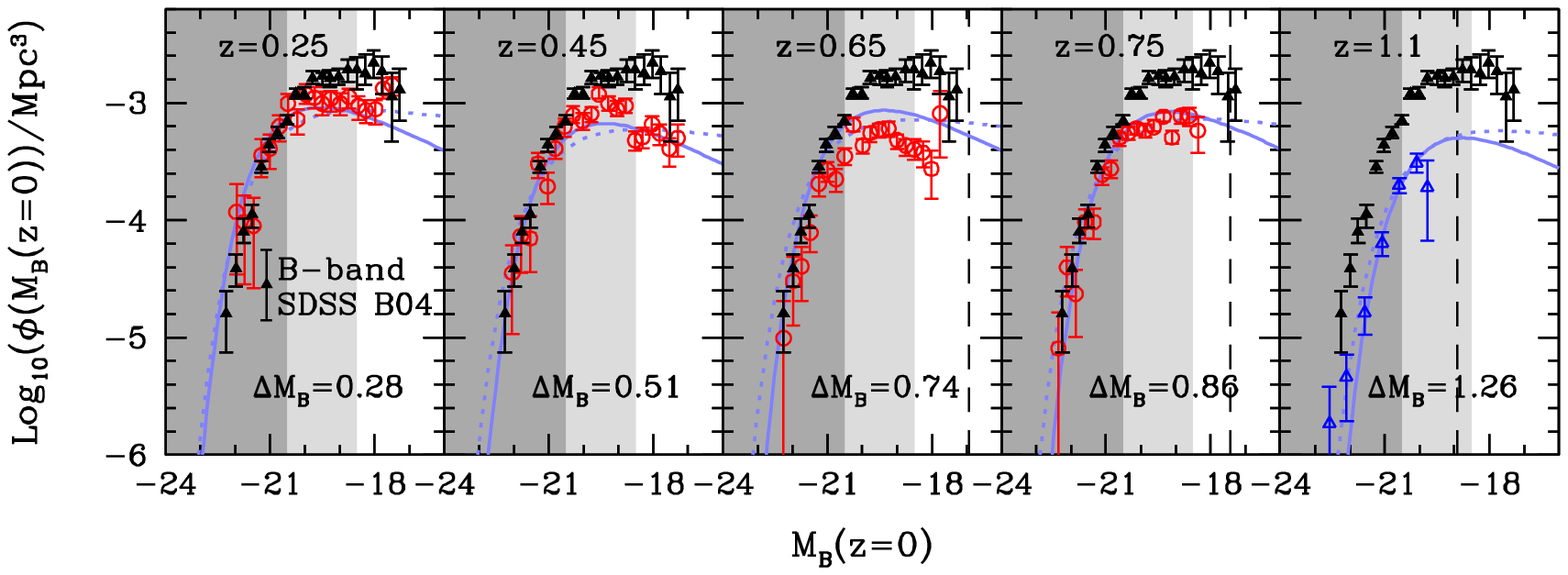}}
\end{picture}
\caption{
Luminosity functions 
of ETGs at different redshifts passively evolved to $z=0$
(with redshift increasing from the left to the right panels as indicated in the key).  
The top and bottom rows show estimates in the r- and B-band, respectively, and for comparison
the black solid triangles show the $z\simeq0.165$ SDSS ETG LF from Benson et al. (2007) in all the top panels, and from
B04 in all the bottom panels.
Red circles show the results from COMBO-17 (B04) and blue open triangles
those from DEEP2 (Faber et al., 2007).
The thick light-blue solid and dotted lines show results from MUSYC, obtained via the PML and
a photo-z based maximum likelihood methods by C09, respectively.
The amount of evolution applied to the $B$ and $r$-band luminosities is indicated in each panel.
The shaded areas delimit the bright and faint ETG populations (dark and light grey, respectively).
The vertical dashed lines (only visible in the subpanels corresponding to $z\geq 0.74$) 
mark the MUSYC completeness limit.
}
\label{fig:lfs}
\end{figure*}

One common method to study the evolution of ETGs
uses PEL function (PELF) or stellar mass function  
measurements of ETGs 
to follow the evolution of the number density of PEL or stellar mass-selected galaxies.  
This method was applied to catalogues
with either photometric redshifts (photo-zs) such as COMBO-17 (Bell et al., 2004, B04)
and the Subaru/XMM-Newton Deep Survey (SXDS, Yamada et al., 2005), or with spectroscopic redshifts 
such as DEEP2 (Faber et al., 2007).  These LFs have been analysed 
by Cimatti, Daddi \& Renzini (2006, CDR) who find that the ETG population is consistent
with no
evolution in their number density since $z\sim 0.8$ (see also Banerji et al., 2010).
We will apply this method to the PELF measurements obtained
from the Multi-wavelength Survey by Yale-Chile (MUSYC,
Gawiser et al., 2006)
by Christlein et al. (2009, C09).
Our approach takes into account the possible
systematic effect imprinted on the luminosity function measurements by errors in
the photo-zs.  This new measurement can
help to study the possible effects from cosmic variance on this particular statistics,
essential to compare with results from galaxy formation models.

The number density evolution of samples of ETGs of equal stellar mass
cannot, in principle, be used to infer a sequence of mergers - or the lack thereof - 
since they do not necessarily constitute causally connected populations.  This
is simply due to the possibility that as time passes, the stellar mass of a galaxy,
even if it is an early-type, can change.
{
For instance, Robaina et al. (2010) use stellar mass-selected samples as the same (statistically)
evolving population, and by measuring the number of close pairs from correlation function estimates,
they infer a small number of $0.7$ mergers per galaxy since $z=1.2$ (see also Patton et al., 2000,
Le Fevre et al., 2000, Lin et al., 2004, Kartaltepe et al., 2007); these estimates could be affected by
biases due to the possible lack of a descendant relationship between samples at different redshifts.
}

The main focus of this paper is to devise and apply a new method of selection of galaxy
samples so that these follow a progenitor-descendant relationship, providing a new, detailed
test for galaxy formation models.
In this first application of the method we combine the LF measurements with 
recent results on the clustering of ETGs from the MUSYC survey (Padilla et al., 2010, P10), which indicate
that the descendants of ETGs are not simply ETGs of similar stellar masses (stellar mass-selected
samples) or evolved luminosities at lower
redshifts.  

{ Our approach will broaden the range of phenomena that can be constrained by
combining LF and clustering measurements since, for instance,
Lee et al. (2009) already used this combination to estimate the star-formation duty cycle
and how its efficiency depends on halo mass for star-forming galaxies at $z=4$ to $6$.}
Using the halo model (see for example Jing, Mo \& B\"orner, 1998, Peacock \& Smith, 2000, Cooray \& Sheth, 2002),
and also combining LF and clustering measurements of red-sequence
galaxies in the Bo\"otes field, Brown et al. (2008, see also Zheng, Coil \& Zehavi, 2005,
White et al., 2007, Wake et al., 2008, Matsuoka et al., 2010) were able to follow the 
growth of the stellar mass of central and satellite galaxies between redshifts $z=1$ and $z=0$
for dark-matter haloes of a given mass.
In particular, our work will be complementary to this 
effort, since we will be able to follow direct descendants while not restricting our analysis
to stellar mass-selected samples.

In the following section we describe the LF measurements we
use in our analysis; in Section 3 we follow the evolution of the number density of 
MUSYC ETG galaxies using PEL-selected samples.  In Section 4
we present the method of clustering selection that allows the construction
of samples that follow a progenitor-to-descendant relationship, and use them to study
the role of mergers in their evolution.
We conclude in Section 5.

\section{ETG passively evolved luminosity functions}
\label{sec:data}

We will analyse the evolution of ETGs using their measured LF
by C09 from  MUSYC.
This survey comprises over
1.2 square degrees of sky imaged to $5\sigma$ AB depths of $U,B,V,R=26$, $I=25$, $z=24$ and $J,K(AB)=22.5$,
with extensive follow-up spectroscopy.  The source detection is done on the combined $BVR$ image
down to a magnitude of $27$ (see Gawiser et al., 2006, for further details).

C09 proposed and applied a new technique,
the photometric maximum likelihood (PML) method,
 to a subset of MUSYC comprising the Extended Chandra Deep Field South (ECDF-S), covering
approximately $0.25$ sq. degrees on the sky.  The PML algorithm 
was used to measure the underlying luminosity function of galaxy populations characterised by
different spectral types.  The latter are 
parameterised with a set of SED
templates from Coleman, Wu \& Weedman
(1980, CWW) or fixed superpositions of two CWW templates, 
extended
into the UV regime using Bruzual \& Charlot (1993) models.  For the present
analysis, we will only use the LF corresponding to the two earliest type templates in this set, which
correspond to an elliptical galaxy and to a E$+20\%$Sbc mix.
P10 demonstrate that the sample of ETGs selected this way is comparable to a selection of the
red-sequence at each redshift (as adopted in e.g. B04, 
CDR, Brown et al., 2008).  They also indicate that these ETGs are compatible with
a sample resulting from a $K<22.5$ selection.

The main advantage of the PML method is that it has been designed to deal with photometric
fluxes alone; it does not require 
estimates of distances or intrinsic galaxy luminosities, and it does
not need to assume a distribution function for redshift errors of any form.
This is achieved by comparing a trial LF to the observed galaxy sample in a parameter space consisting of
observed fluxes, whose error function is well understood.
Therefore, the PML method explicitly avoids a second convolution with the photo-z
errors, being in principle free of this possible source of systematic effects.

For comparison purposes, C09 applied a maximum likelihood method to provide estimates of photo-zs
with competitive accuracies with respect to other photo-z measurement algorithms
such as BPZ (Benitez, 2000).  In the case of MUSYC, the resulting redshift
uncertainty is $\Delta z\simeq 0.065 \times (1+z)$ (normalised median absolute deviation, see 
Hoaglin, Mostelen \& Tukey  1983).
C09 also used these estimates of photo-zs to measure the LF of ETGs using a classic
maximum likelihood method.  We will compare the results on ETG evolution using
both C09 LF estimates.

We use the rest-frame r- and B-band LF estimates from MUSYC ETGs to allow a direct
comparison to LF measurements in both bands.  All the ETG LFs are passively evolved 
down to $z=0$ by applying 
an empirical passive evolution recipe whereby the evolved luminosity can be found via
\begin{equation}
M_B(z=0)=M_B(z)+1.15z,
\label{eq:ev}
\end{equation}
following results from van Dokkum \& Stanford (2003),
Treu et al. (2005), and di Serego Alighieri et al. (2005).  
P10 showed that this passive evolution recipe is well followed by a $[Fe/H]=0.3$ single stellar
population (SSP) which at $z=1$ is $3.5$Gyr old,  and used this SSP to
work out the equivalent recipe for the r-band, which is fit by
\begin{equation}
M_r(z=0)=M_r(z)+0.98z.
\label{eq:evr}
\end{equation}
This SSP is characterised by a colour which varies slowly with redshift,
\begin{equation}
B-r\simeq1.32-0.17\times z.
\label{eq:color}
\end{equation}

We apply these evolution corrections to B- and r-band LF measurements from
relatively high redshift samples and
compare them to results at $z\simeq0.165$ from
the Sloan Digital Sky Survey (SDSS; York et al., 2000) for ETGs by B04 in the B-band 
(selected using the red-sequence), and in the
r-band by 
Benson et al. (2007, Be07), who selected ETGs as sources with a dominant bulge component
(an alternative ETG LF measurement from SDSS is provided by
Ball et al., 2006).   
{ We find that the two estimates 
are perfectly consistent with one another when applying Eq. \ref{eq:color}.}
When correcting the luminosities by passive evolution and using this
quantity to select samples of ETGs (with little or no star formation), we effectively mimic 
a stellar mass selection of galaxies (CDR).  

The comparison between the ETG LFs at different redshifts is shown in Figure \ref{fig:lfs}, where 
the Be07 and B04 SDSS ETG LFs are plotted in all the top and bottom panels, respectively.
The MUSYC PML and classic maximum likelihood measurements of the ETG LFs for different redshift ranges 
(indicated in the panels) are shown for the r- and B-band 
(top and bottom panels, respectively).
The bottom panels also show the LF estimates
from COMBO-17 and DEEP2 
for the same redshifts.
As can be seen, the bright ETG population from COMBO17
matches the ETG SDSS result for all the
redshift slices shown in the figure.  
Only the high redshift DEEP2 data shows a lower comoving number density than
the SDSS LF at this luminosity range.  Therefore, from these results one would
conclude that there is no evolution in the number density of evolved ETGs from
$z\sim0.85$ onwards.  \footnote{
Notice that this does not prove there are no mergers since 
it would still be possible that galaxies of a given luminosity merge to
form brighter galaxies while maintaining a close to constant luminosity function (see Section 2)
with galaxies in this bin leaving to higher luminosity bins.
}
The MUSYC results (light-blue lines) for the bright population, as obtained from the
PML algorithm or even from the classic maximum likelihood method, on the other hand, 
show indications of some evolution
{ in the number density of bright ETGs} with redshift { in both bands},
even at $z<0.45$.  { This is more clearly seen in the r-band, 
than in the B-band when comparing to the
$z=0$ SDSS LF since, in the latter case, the SDSS ETG LF characteristic luminosity
appears to be slightly fainter than for the low redshift MUSYC ETG, even when the B- and r-band
SDSS LFs are entirely consistent when taking into account Eq. \ref{eq:color}.  This could
indicate that the MUSYC ETG samples at low redshift are slightly redder than
the SDSS ETGs.}
We will leave the discussion of the faint
population to the following section.

\section{Evolved ETG luminosity selection of samples in MUSYC}

We study the number density evolution of PEL-selected samples
from MUSYC, and compare it to previous results from
 DEEP2, COMBO-17 and SXDS. 

{ This approach has been applied in several ocasions in the literature, and we apply
it in this section to MUSYC ETG galaxies.  This procedure
can only be used to follow the evolution of the number density
of ETGs of a particular PEL (or stellar mass).
Due to their simplicity, selections of this type have been assumed to provide samples of the
same evolving population at different redshifts (e.g. CDR, Robaina et al., 2010, Marchesini et al., 2009).  However,
they cannot be used to study the
frequency of mergers in principle, since the stellar mass or evolved luminosity
of an early-type galaxy can evolve with time.  i.e., 
there is no precise information on whether the two populations
are connected in a progenitor-descendant relationship (P10).

Notice, however, that this approach still allows a comparison with models,
and in particular it has been used 
in galaxy formation models (De Lucia et al., 2006, Benson \& Devereux, 2010) to study the
number density of ETGs of similar PEL and similar stellar masses, respectively.
In this section we will integrate the ETG LFs over two PEL ranges (shown
by the dark and light shaded areas in Figure \ref{fig:lfs}) delimited by
$M_B(z=0)<-20.5$, for the bright ETG population, and by $-20.5<M_B(z=0)<-18.5$
for the faint ETG population to obtain their number densities $n_{bright}(z)$ and $n_{faint}(z)$.
The r-band limits can be obtained using $B-r=1.15$ (Eq. \ref{eq:color}).
The luminosity $M_B(z=0)=-20.5$ corresponds
approximately to a stellar mass of $10^{11}M_{\odot}$ (CDR), a limit used
by De Lucia et al., in their analysis of semi-analytic galaxies.
In this paper, we evolve the r-band 
$z=0.165$ descendant luminosities estimated by P10 to $z=0$
using Eq. \ref{eq:evr}.
By integrating over luminosity the method is less
sensitive to the exact functional form adopted for the LF.
Our analysis of mock catalogues in Appendix A2 ensures little influence from
systematics coming from the LF measurement method or photo-z errors.

Our measurements of number density evolution from MUSYC galaxies will be compared to those
already published for
the datasets COMBO-17, DEEP2 and SXDS, and the De Lucia et al. (2006) models of galaxy formation.
To study the evolution of the number density of ETGs 
we need $z=0$ comparison samples for which
we use the $z=0$ ETG LF from the SDSS measured 
by B04 for the B-band, and by Be07 for the r-band.
Figure \ref{fig:ratio} shows the resulting bright and faint ETG number density evolution
(top and bottom panels).  As can be seen, the number
density of bright ETGs (top panel) for COMBO-17, DEEP2 and SXDS is roughly consistent with that of 
the present-day bright ETGs up to redshifts
$z\sim0.7$, indicative of little evolution since then (CDR).
The dashed lines correspond to the evolution of model ETG number densities inferred 
by De Lucia
et al. (2006) which, for the bright ETGs, lie below the observational points for most
of the explored redshift range; at low redshift where the volume of the samples
is small, there is a marginal agreement between the De Lucia et al. results and the observations
($z<0.35$).

\begin{figure}
\begin{picture}(250,180)
\put(-0,-115){\includegraphics[width=11.cm,clip]{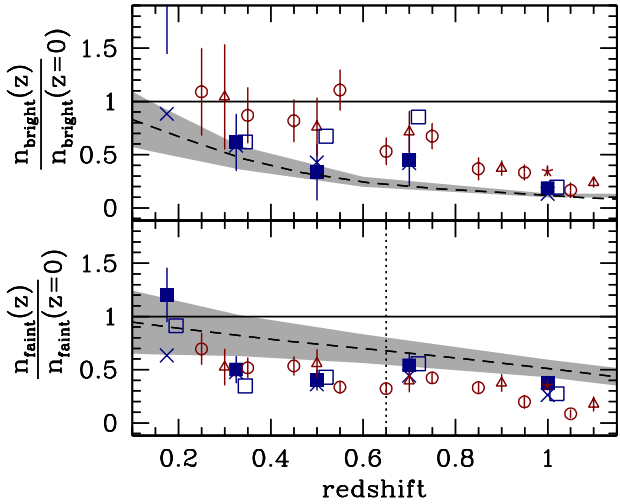}}
\end{picture}
\vskip .4cm
\caption{
Ratios of ETG number densities at a given redshift
with respect to a $z\sim 0$ SDSS sample.   
The horizontal solid lines show the unit ratio.  The shaded areas show the cosmic variance
as calculated in Appendix A2.
The squares show the results from MUSYC for the r-band (solid symbols) and the B-band (open symbols,
slightly displaced on the x-axis) from PML estimates of the LF;
the errorbars for the results in the two bands (r and B) are similar, but 
are shown only for the r-band to improve clarity.  The crosses
show the results from the r-band LF calculated using the photo-z based maximum likelihood method. The 
circles correspond to results from the COMBO-17 ETG LF estimate, triangles 
to DEEP2, and stars to the SXDS.
Top panel: samples of bright ETG galaxies. Lower panels: faint ETGs.
Comparing the shaded area and symbols, a marginal agreement between the evolution in models and observations can be noticed.
}
\label{fig:ratio}
\end{figure}

The squares (solid correspond to the r-band, open to the B-band) 
in this figure show our results for the MUSYC ETG LFs, obtained via the PML
method.  As can be seen,
the PML measurements in the B-band show a good agreement with all the other observational results
presented here from DEEP2, COMBO-17 and SXDS.  The results from both bands, and in
particular from the r-band, also show a very good 
agreement with the De Lucia et al. (2006) semi-analytic model results for the evolution of the
bright ETGs.  The agreement for the different bands varies slightly probably due { to 
the slightly redder colours of the MUSYC ETGs in comparison to those from SDSS.}

As can be seen, to the level of certainty allowed by the combination of the different
observational estimates, 
it is difficult to ascertain whether the observed evolution of the bright ETG
number density is consistent with this particular model.  
Notice, however, that results from { some of the catalogues shown here taken individually,} could be used to
rule out this particular model; this shows the importance of using different pencil-beam
directions to be able to account for the important cosmic variance in the counts.
This also affects to some degree 
the conclusions from Hopkins et al. (2008), 
since the observational results from $z\leq1$ only 
provide loose constraints to disentangle between different star-formation quenching
mechanisms.  

The evolution of the faint ETG number density shows consistent results between 
the different catalogues,
and are slightly below the semi-analytic estimates, even at redshifts below
$z=0.65$ where the recovery of the number densities in the mocks was succesful (See Appendix A2).  
However, at least for MUSYC galaxies,
the expected cosmic variance (See Appendix A2) is able to account for some of these differences.

\section{Clustering selection of descendant samples}

In this section we present a new method to select samples of galaxies at different redshifts 
that follow a progenitor-to-descendant relationship in a statistical way.  Such samples
can be used to constrain the role of mergers
in the evolution of ETGs by comparing their number densities to that of their inferred
descendants.   This method can also be applied to samples of galaxies with available
stellar masses, instead of PELs.    

Instead of adopting a stellar mass selection, the clustering-selected
descendants are obtained by taking a sample of haloes from the simulation with the dark-matter
masses corresponding to the hosts of ETG galaxies at a given redshift obtained via clustering 
measurements (P10).
Then these haloes are followed to $z=0$ using merger trees extracted from
the numerical simulation, and their clustering is compared to that of ETGs in the SDSS 
(z=0, from Swanson et al., 2008).  The SDSS ETGs with this clustering are then selected as
the descendants of the MUSYC ETGs.  
P10 used this method to connect different ETG populations at $0.25<z<1.4$ to 
ETG galaxies at $z<0.25$. 
This study shows that in a $\Lambda$CDM cosmology it is not correct to consider
ETGs with equivalent passively evolved $z=0$ luminosities at different redshifts
as descendants/progenitors of one another since the low redshift galaxies with
the same PEL/stellar mass show a lower clustering than is expected for the
descendants.  

P10 find that MUSYC ETGs with passively evolved $z=0$ luminosities
$M_r(z=0)<-19.7$, evolve from typical luminosities, { expressed in units of the ETG $L^*$ }
of $L/L^*(z)\simeq 0.7$ at $z=1.15$ into
galaxies with typical $L/L^*(0) \sim 2.1-5.2$ at $z=0$, and from $L/L^*(z)\simeq 1$ galaxies
at $z=0.35$ to $L/L^*(0)\sim 0.35-0.8$.\footnote{The ranges in descendant $L/L^*(0)$ correspond to 
variations in the results between the EHDF-S and ECDF-S fields.}
{ Table \ref{table} shows the typical descendant $z=0$ B-band magnitude and stellar mass for our
samples of progenitor galaxies.  Notice that as progenitors are selected using the same PEL (or stellar
mass) cut, they are {\it not} expected to lie on the same evolutionary line; this is confirmed by the
different typical descendant properties shown in the table.}

\begin{table}
\caption{Properties of clustering-selected $z=0$ descendants of $M_r(z=0)<-19.7$ (stellar masses above 
$\sim 10^{10}$h$^{-1}M_{\odot}$) progenitors at different redshifts.}
\label{table}      
\centering                          
\begin{tabular}{c c c}        
\hline\hline                 
Progenitor & Descendant & Descendant \\    
Redshift & $M_r(z=0)$ & $\log_{10}(M_s/$h$^{-1}M_{\odot})$ \\
\hline                        
& & \\
   0.175 & $-17.5^{-0.7}_{+3.7}$ & $9.4^{+0.4}_{-2.0}$ \\      
   0.325 & $-18.8^{-0.4}_{+0.6}$ & $10.1^{+0.2}_{-0.3}$ \\
   0.5   & $-19.4^{-0.3}_{+0.4}$ & $10.4^{+0.2}_{-0.2}$ \\
   0.7   & $-20.2^{-0.3}_{+0.5}$ & $10.9^{+0.2}_{-0.3}$ \\
   1.0   & $-20.8^{-0.4}_{+0.6}$ & $11.2^{+0.2}_{-0.3}$ \\ 
& & \\
\hline                                   
\end{tabular}
\end{table}

We calculate the number density of descendants using the B04 and Be07 SDSS ETG LFs for
galaxies with median luminosities corresponding to the descendants of a given sample
of MUSYC ETGs at redshift $z$.
The number density of progenitors is calculated at redshift $z$, using a lower limit in
$M_r(z=0)=-19.7$, or equivalently, $M_B(z=0)=-18.55$, consistent with the lower limits
used in P10; { according to the relation between $z=0$ B-band luminosity and stellar mass by
di Serego Alighieri et al. (2005, Figure 13 in their paper), 
this luminosity cut effectively selects ETG galaxies with stellar masses $M_s>10^{10}$h$^{-1}M_{\odot}$.
The ratios between the number density of progenitors and descendants} 
are shown as squares in the upper panel of Figure \ref{fig:ratiodesc} (open 
symbols for the B-band, filled symbols for the r-band).  
The errorbars in this panel correspond to the uncertainties
in the descendant luminosity, extracted from P10.
As can be seen,
the ratio is significantly higher than unity at $z>0.6$ in both bands, indicating the need for mergers
between ETGs in order to diminish their number density towards $z=0$.\footnote{
{ The mock catalogue analysis in Appendix A shows a possible overestimate of a factor $1.5$ at this redshift
range, which we will bear in mind in our conclusions.}
}  Lower redshift samples
show number densities similar to their expected descendants.
As can be seen, the photo-z based maximum likelihood method (crosses, shown only for the r-band 
for clarity) provides 
results in agreement with those from the PML based LFs.

\begin{figure}
\begin{picture}(250,200)
\put(-5,-20){\includegraphics[width=11.cm,clip]{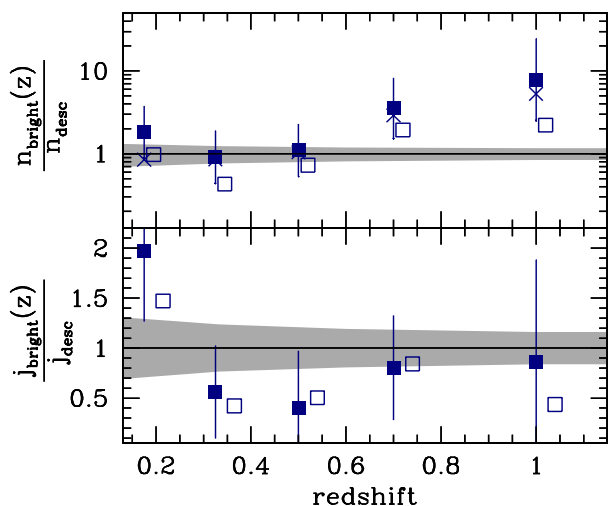}}
\end{picture}
\caption{
Top panel: Ratio between the number density of progenitor ETGs 
with stellar masses $M_s>10^{10}$h$^{-1}M_{\odot}$
at a given redshift and their expected $z=0$
descendants, as inferred from
the clustering analysis of P10.  The filled squares show the results from
using the PML r-band LF estimates; open squares show the
results for the B-band.  Errorbars are only shown for the r-band result to improve clarity;
errors for the B-band show similar amplitudes.
The crosses correspond to the results from the photo-z based maximum likelihood
method to calculate the LF in the r-band.
The solid line shows the unit ratio and 
the grey shaded area shows the estimated cosmic variance in a $0.25$sq. degree light-cone survey divided
in slices of $\Delta z=0.1$.
As can be seen, the number density of high redshift progenitors is higher than that of their $z=0$ descendants; this
is not the case for progenitors at $z<0.6$ and their $z=0$ descendants.
{ Bottom panel: ratio between the luminosity density of progenitor ETGs and that of their $z=0$ ETG descendants 
in the r- and B-bands (filled and open squares, respectively).  Regardless of the progenitor redshift, the luminosity
density ratios shown are consistent with the unit value.}
}
\label{fig:ratiodesc}
\end{figure}

Taking advantage of the measured LFs for the high-redshift ETG samples, we calculate
the ratios between the luminosity densities of the high-z ETGs and that 
of their $z=0$ likely descendants, using the PML LF measurements.  
This is shown in the lower panel of Figure \ref{fig:ratiodesc} as squares
(solid symbols
correspond to results in the r-band, the open symbols to the B-band). 
As can be seen,
regardless of the photometric band
the data shows that 
as the redshift decreases the luminosity density of ETGs is consistent with the unit value
(bearing in mind that our different high redshift ETG samples are {\it not} in the same
evolutionary line).  The lowest redshift point is marginally above the unit ratio, but is
also affected by the largest poisson uncertainties in the LF measurement due to the small number
of galaxies at such low redshifts in the ECDF-S, and also by the large effect from cosmic
variance due to the small volume.

The results from this analysis show that ETGs at $z=1$ are likely to descend
into $z=0$ ETGs undergoing a decrease in space-density
of a factor $5.5\pm4.0$ (combining the results from the B and r bands).
But, due to the constant luminosity density, mergers would provide the
required increase in luminosity without the need for important episodes of star-formation.  
Notice that the amount of mergers derived here (of about four mergers since $z=1$ for each $z=0$
galaxy) is higher than that estimated
from close pairs of stellar mass-selected samples at different redshifts; Robaina et al. (2010)
find that galaxies undergo $\sim 0.7$ mergers between $z=1.2$ and $0$.  We suggest that a more consistent measurement
of merger rates using close pairs would require the use of samples in a more likely progenitor-descendant
relationship than that provided by a stellar mass selection.
The use of clustering-selected samples in this case would include brighter low redshift
samples which could change the resulting merger rates.

It should be borne in mind that our analysis does not consider sources or sinks, { but since
a quantitative analysis of their influence requires the use of models that include
the evolution of dark-matter haloes in a fully non-linear way (merger trees) including
galaxies, we defer this to a forthcoming paper (Padilla et al., in preparation).
In this paper we present possible sources of these systematic effects, and in some
cases, qualitative estimates.}

On the one hand, we have not taken into account the possibility that some
of the stellar content of ETGs may be lost, for instance into the intracluster medium (ICM) due
to tydal stripping,
before a merger takes place.
Conroy, Wechsler \& Kravtsov (2007) showed that in order to reproduce the 
evolution of the massive end of the stellar mass function since $z=1$, the $z\sim0$ brightest cluster
galaxy-cluster mass relation, and the intracluster light luminosity distribution, up to $80$ percent of
the stellar content of galaxies in disrupted substructures (i.e. only a fraction of the satellites) 
within dark-matter haloes should be lost to the intracluster medium; { in the next Subsection the analysis
of the halo model shows some evidence for sinks consistent with this estimate}.  
Additionally, the star-formation activity could be reignited via new cold gas infall or AGN shutdown, which could
also remove the galaxy from the ETG sample.
On the other hand, blue galaxies could become ETGs and contaminate the descendant population.

{
Notice that the approach followed by P10 to obtain a typical host DM halo mass 
does not need information on the stellar masses of galaxies, only on their clustering.  However
it is necessary to take into account the clustering amplitude vs. stellar mass relation 
since the P10 approach to find descendants
requires a one-to-one relation
between PEL or stellar mass and host halo mass.  
In the case
of $z=1$ MUSYC ETGs this relation shows a minimum at $M_B(z=0)=-18.5$ (P10)
which indicates that our approach can only be applied to galaxies brighter than this lower limit.
In order to follow the
descendants of fainter galaxies it would be necessary to adopt, for instance,
a subhalo abundance matching method.  The latter
provides a direct connection between DM subhalo mass (at
infall) and stellar mass (e.g. Kravtsov et al., 2004), but 
is subject to uncertainties even when clustering and mass function measurements are available
(Neistein et al., 2001). 
 } 

\subsection{Clustering-selected samples and halo model analyses}

Further clues on the evolution of ETGs were found by
Brown et al. (2008) who analysed the Bo\"otes field.  Using the Halo Model they point out 
that central ETG galaxies at $z=0$, which occupy the bright-end of the LF,
have only increased in about $30$ percent their stellar mass since $z=1$, whereas 
the stellar masses of satellite
galaxies have increased by a factor of $\sim 3$.  

Applying their universal Halo Occupation Model fits (equations $12$ and $13$ in their paper) to our results
for the $z=1$ MUSYC and their descendant $z=0$ SDSS ETG populations
we find that in median, 
(i) $\sim80\pm5$ and $\sim 94\pm3$ percent of the progenitors\footnote{
We apply the HOD to the unevolved luminosities of the progenitors
as its parameters are provided for rest-frame ETG magnitudes by Brown et al.
} and descendants (respectively) are central
galaxies.  (ii) The latter allows us to separate centrals and satellites by sharp cutoffs in luminosity 
as a first approximation, with which we obtain that
centrals increase their luminosity by a factor of $1.7^{+2.2}_{-0.5}$, whereas satellites do so
by a factor of $2.5^{+1}_{-1.2}$ (consistent with the Brown et al. estimates),
(iii) and that the number density of centrals and satellites decreases
by factors of $\sim 4\pm2$ and $\sim 10\pm7$, respectively. { Taking into account the relative
numbers of centrals and satellites, this is in agreement with the overall factor of $5.5\pm4.0$
decrease in number density of ETGs}.

{
The percentage of galaxies which underwent major mergers has been predicted by
semi-analytic models of galaxy formation.  Kochfar \& Silk (2009) show that 
between $1\%$ and $2\%$ of the massive galaxies (with stellar masses similar to
those in our samples) undergo a major merger in their last Gyr for $z<=1$.

Our results are consistent with the Kochfar \& Silk estimates.
The decrease in the descendant
MUSYC ETG number density can be the product of both minor and major mergers (considering
the latter those with a mass ratio of $1/3$ or higher, e.g. Lagos, Cora \& Padilla, 2008).
Taking into account the shape of the progenitor ETG luminosity function at $z=1$ and
separating centrals and satellites by sharp luminosity cuts, we make a Monte-Carlo estimate
of the ratios between luminosities merging galaxies.   Our
average result of a factor of $5.5\pm4.0$ decrease in the number density of ETGs points to about
four mergers per galaxy, where  
we assume that each one increases the brightness of the central galaxy.  The ratios between
merging galaxies are shown in Figure \ref{fig:mergers}, where the shaded region corresponds
to major mergers and as can be seen, only a small fraction of the total merger events (solid line)
fall into this category.
We find that $\sim 31\%$ of the central galaxies
experience a major merger since $z=1$, with a decreasing frequency towards $z=0$; this translates
into an average $\sim4\%$ of ETGs undergoing a major merger in their last Gyr.  
We also find that $2/3$ of the major mergers involve a $z=1$ central galaxy that became a satellite and then
merged with a larger central galaxy.

\begin{figure}
\begin{picture}(230,230)
\put(-5,-10){\includegraphics[width=8.6cm,clip]{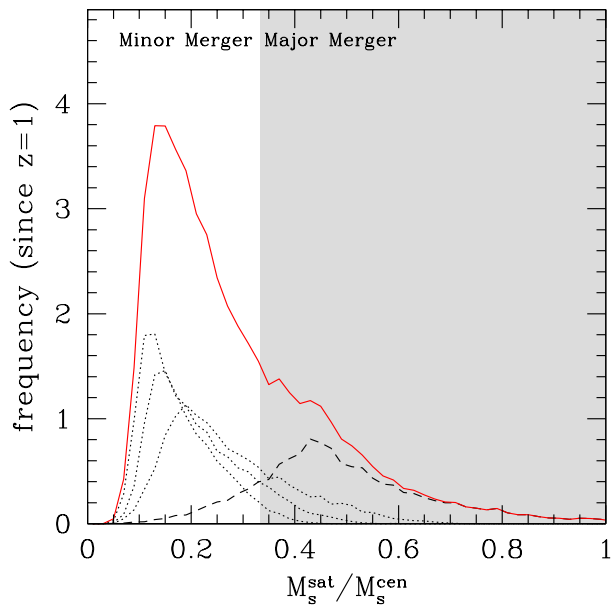}}
\end{picture}
\caption{
Frequency of stellar mass ratios of ETG galaxies involved in mergers, for the case of
$z=1$ progenitors with $M_r(z=0)<-19.7$ (i.e. stellar masses above $\sim 10^{10}$h$^{-1}M_{\odot}$) 
and their $z=0$ descendants for a total of $4$ mergers per descendant.  The solid line shows the sum
of all the merger events, dashed corresponds to mergers between central galaxies, dotted to mergers
between a central and one of its satellites.  The shaded region corresponds to ratios above the limit for a major merger (mass ratio of $1/3$ and higher).
}
\label{fig:mergers}
\end{figure}

This simple calculation ignores possible correlations between central and satellite
galaxy luminosities and also the evolution of the ETG LF between $z=1$ and $0$.  Therefore,
these results should only be taken as 
a rough estimate of the frequency of major mergers.  A more thorough analysis
would require the use of conditional luminosity functions of satellite luminosities in terms of
those of their central galaxies, on the one hand, and more precise measurements of LF and clustering
of ETGs in this range of redshift, on the other.
An alternative approach is to take into account
merger trees of dark-matter haloes and to follow the evolution of galaxies within them.  We will perform this
analysis in a forthcoming paper where the dark-matter haloes will be selected so as to mimic
the high-z MUSYC ETG population (Padilla et al., in preparation).
}

The Halo Model analysis also allows us to 
infer the total luminosity in progenitors (adding centrals and satellites together) which exceeds that of
descendants by a factor of $\sim 4^{+4}_{-2}$, which points to a possible sink in the 
luminosity of the ETG population { of similar amplitude to the expected effect of the ICM pointed
out by Conroy, Wechsler \& Kravtsov (2007)}.  Additionally,
the hint that the decrease in the number density of satellites is larger than that of centrals would
indicate that a majority of the mergers would occur with central
galaxies of { smaller mass, progenitor dark-matter haloes} before joining the final, more 
massive halo\footnote{
{ The reason for this is that if there were no galaxy mergers in the progenitor haloes, 
most of the $z=1$ satellites would survive until today due to the fact that the
dynamical friction timescales are much larger for lower satellite-to-central mass ratios} 
(see for instance Boylan-Kolchin et al., 2008).}
consistent with independent indications pointing to the same conclusion 
(Gonz\'alez \& Padilla, 2009, Porter et al., 2008).

\section{Conclusions}

In this paper we presented a new method to select
ETG samples that follow a progenitor-to-descendant relationship which uses clustering and luminosity
or stellar mass function information.  We used this method to define clustering-selected
ETG samples from the MUSYC ECDF-S field, which compared to their clustering-selected
descendants in the SDSS allowed us to study the importance of mergers in their
evolution.


Our main results and conclusions can be summarized as follows.
\begin{itemize}
\item[(i)] 
 From the analysis of the number density of samples of MUSYC 
ETGs with equal luminosities, passively
evolved to $z=0$ (a procedure that effectively compares samples of galaxies with similar stellar
masses), we find 
an evolution of the abundance of bright ETGs 
characterised by a steady increase even down to $z\sim0.3$, 
consistent within the errorbars with results from COMBO-17, DEEP2, and
the De Lucia et al. (2006) measurement using semi-analytic galaxies.
COMBO-17 and DEEP2 show
slightly less evolution in the ETG number density and a lesser agreement with the model,
showing the importance of numerous pencil-beam like catalogues to pin down cosmic variance.
We point out that, in principle, 
this selection does not produce samples obeying a progenitor-to-descendant relationship (as
assumed in several works, e.g. CDR, Robaina et al., 2010) even though it 
can still be used for meaningful comparisons
between models and observations.

\item[(ii)]  { We  implemented the clustering selection method with the aim of finding 
clues on the role of mergers in the evolution
of ETGs.  
We define progenitor ETG samples from MUSYC with median redshifts between $0.2$ and $1$.
Their $z\simeq0.1$ descendant galaxies are selected so as to have the same clustering
amplitude
as the dark-matter haloes that host the high-z ETG samples (obtained from clustering measurements by P10)
followed down to $z=0$ in numerical simulations.}  
The analysis of these samples allowed us to find some evidence for dry mergers of ETG galaxies between
redshifts $z\sim 0.8$ and $z\sim0.2$, 
since their number density declines by a factor $\sim 5.5\pm4.0$
(Figure \ref{fig:ratiodesc}, top panel, the ranges show the variation in the
results from using the B- and r-bands).  { Possible systematics in the LF measurement
obtained in our analysis of mock catalogs
could lower this factor to $\sim 3.8\pm2.6$.}  Given
that the luminosity density of progenitors and descendants are consistent with one another
(Figure \ref{fig:ratiodesc}, bottom panel),
 our results point towards dry mergers without important star-formation
episodes in ETGs, at least down to the precision allowed by the MUSYC dataset.   
These two results are consistent in the sense that 
{ at $z=0$  there are not enough ETGs to
account for the progenitor population} via simple passive evolution; the merging scenario helps solve
this problem, by decreasing the number density of fainter galaxies, which after merging provide the
observed ETG luminosity and number density at $z=0$.  

\item[(iii)]
Combining our results for the descendants of $z=1$ MUSYC ETGs
with halo occupation fits to LF and clustering measurements by Brown et al. (2008),
{ we find that $\sim 4\%$ of the descendants of $z=1$ ETG galaxies with stellar masses 
above $\sim10^{10}$h$^{-1}M_{\odot}$ undergo a major merger in their last Gyr, in agreement with recent
results from semi-analytic models by Kochfar \& Silk (2009).  We also find that two thirds of the
major mergers
involve an ETG which was central galaxy at $z=1$, and some }
indications that most of the ETG mergers would need to take place in groups 
before these fall into large clusters (in agreement with 
Porter et al., 2008, and Gonz\'alez \& Padilla, 2009).
\end{itemize}

In the analysis of (ii) and (iii)
we have assumed that the $z=0$ ETG sample with the same clustering as the
descendants of the haloes hosting high redshift ETGs does not include galaxies which turned
red and dead in between.  
These would contribute to the number density of ETGs
at $z=0$,  which would then be overestimating that of the true descendants of the individual
ETG progenitors.  On the other hand, our approach has not considered sinks for the ETG population,
which could be provided by tidal effects in the intracluster medium or by processes that reignite
the star-formation activity in galaxies, which in turn would overestimate the number density of
the progenitors.  

Upcoming surveys such as those planned for the Large Synoptic Survey Telescope (Abell et al.,
The LSST Science Book, 2009), will
allow much better statistics by increasing the solid angle with respect to the currently available
deep photometric surveys such as MUSYC and COMBO17.  Furthermore, with a higher signal it could also
be possible to produce measurements of the expected merger rates and star formation history of
combined samples of early and late type galaxies, and via comparisons with models of galaxy formation
to help improve our understanding of the evolution of galaxies from high redshifts to the present day.

\begin{acknowledgements}
We thank Carlton Baugh, Cedric Lacey, and the anonymous Referee for helpful comments and suggestions.
NDP was supported by a Proyecto Fondecyt Regular
no. 1110328.  This work was supported in part by the "Centro de Astrof\'\i sica
FONDAP" $15010003$, by project Basal PFB0609, and
EG was supported by the National Science Foundation
under Grant. No. AST-0807570.
\end{acknowledgements}

\begin{appendix}
\section{Mock catalogue analysis}
\label{sec:corr}

We use the ECDF-S mock galaxy catalogues from C09 (two in total)
to analyse the effects of distance uncertainties on the LF.
The mocks are
 extracted from a 
$\Lambda$CDM numerical simulation
populated with {\small GALFORM} (version corresponding to Baugh et al., 2005)
semi-analytic galaxies. 
The simulation contains $10^9$ DM particles in
a periodic box of $1000$h$^{-1}$Mpc a side. The cosmological
model adopted in this simulation is characterised by a
matter density parameter $\Omega_m=0.25$,
a vacuum density parameter $\Omega_{\Lambda}=0.75$, a
Hubble constant, $H=h 100$kms$^{-1}$Mpc$^{-1}$, with $h=0.7$,
and a primordial power spectrum slope, $n_s=0.97$.  The
present day amplitude of fluctuations in spheres of
$8$h$^{-1}$Mpc is set to $\sigma_8=0.8$.  This particular
cosmology is in line with recent cosmic microwave background
anisotropy and large scale structure 
measurements (WMAP team, Dunkley et al., 2009, Spergel et al. 2007, see also S\'anchez et al., 2006).  
We adopt this cosmology for all the calculations performed throughout this paper.

In order to make the mocks as similar to the observations as possible, C09 used the information on
stellar mass and bulge fractions to assign a SED drawn from
a smooth continuum of possible templates (from the same set used to calculate photo-zs) 
to each galaxy. This SED is used to calculate
fluxes in the MUSYC bands.  Using these photometric data,
C09 repeated the photo-z measurement procedure on the mock galaxies, which are
affected by similar $z_{\rm phot}$-$z_{\rm spec}$ error distributions as the actual MUSYC data, including
outliers and catastrophic errors.

The two mock catalogues match the angular selection function of MUSYC fields but
{ include no evolution since they} were constructed using only the $z=0$ simulation output. 
In order to mimic an {\it active} evolution of the mock ETG galaxy luminosities with redshift, we apply 
a linear magnitude brightening of
$0.6$ magnitudes between $z=1$ and $z=0$.  This choice produces an evolution of the number density
of bright ETGs consistent with the reported evolution
of ETGs in the De Lucia et al. (2006) model {(also consistent with the evolution of
{\small GALFORM} ellipticals found by Benson \& Devereux, 2010)}.  We apply this brightening
of the LF to our mock results from this point on.

\begin{figure}
\begin{picture}(240,230)
\put(5,-20){\includegraphics[width=9.cm,clip]{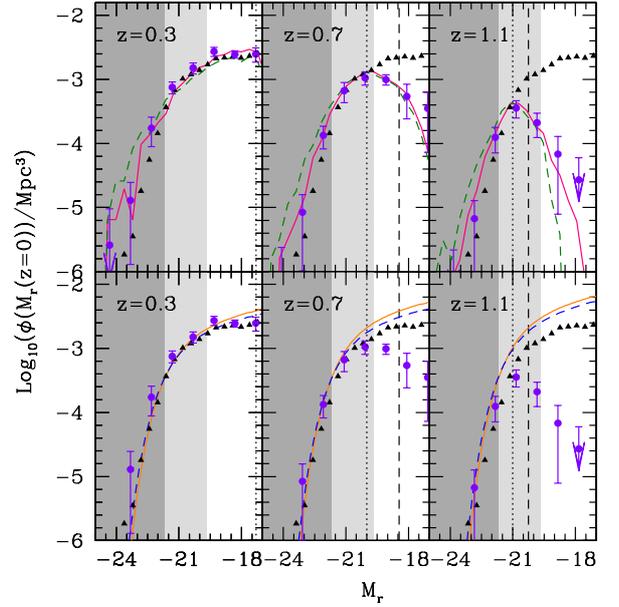}}
\end{picture}
\caption{
Effect of redshift measurement errors on 
the luminosity function as found using mock MUSYC catalogues, analysed in
three redshift bins (increasing from left to right).
In all the panels, the triangles show the true, evolving r-band ETG LF of the full simulation box.
Top panels: LFs calculated using the $1/V_{max}$ weighting method.
Filled circles show the results from using the true redshifts in the mock catalogues,
the solid lines indicate the
results when adopting photo-zs, the dashed lines shows the effect
of a further convolution with the photo-z errors. 
Bottom panels: the filled circles are as in the top panels.  The solid lines correspond
to the PML estimates of the LF from the mock, the dashed lines to the photo-z based
maximum likelihood estimator (C09).
Shaded areas and vertical dashed lines are as in Figure \ref{fig:lfs}.  { The vertical
dotted line shows the magnitude above which the recovery of the LF using the true redshift
starts to degrade.  Errorbars show poisson uncertainties.  }
}
\label{fig:mocks}
\end{figure}

\begin{figure}
\begin{picture}(180,175)
\put(30,-10){\includegraphics[width=6.5cm,clip]{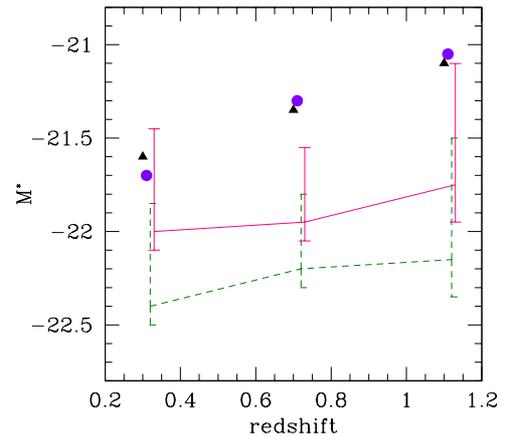}}
\end{picture}
\caption{
Recovered values of $M^*$ from the LFs in the top panels of Figure \ref{fig:mocks}, 
as a function of redshift.  The results from fitting the true LF are shown as triangles, from the
true redshifts in the mock as circles, from the photo-zs as solid lines, and from the double convolution
as dashed lines.  Errorbars show the $1-\sigma$ confidence ranges obtained from the fits.  Small
displacements in redshift were applied to improve clarity.
}
\label{fig:mstarmocks}
\end{figure}

{ In this appendix we use six extimates of the LF.  Those obtained using
the LF of ETG galaxies in the full simulation box, 
the PML and classical photo-z based maximum likelihood estimates, and
$1/V_{\rm max}$ estimates of the LF from the mocks using 
the true galaxy redshifts (with no errors), the photometric redshifts, and
the latter with a further convolution with the photo-z errors.}

Figure \ref{fig:mocks} shows the LF of the combined set of galaxies
from both mock catalogues; this is done in order to minimise random errors and focus our analysis
on possible systematic effects.  From left to right, the panels show different
redshift slices
selected by placing limits on photo-zs (the average
redshifts are indicated in each panel).
The triangles represent the true, underlying luminosity functions in the simulation, and circles
show the $1/V_{max}$ estimate of the LFs from the mocks using the true galaxy redshifts. 

In the top panels the lines correspond
to LFs calculated using the $1/V_{max}$ weighting with photo-zs.
The solid line shows the resulting LF from assuming the
photo-z provides the correct distance to the galaxy; this approach
produces an excess of bright galaxies (with respect to the result from the true redshifts) 
which is slightly more important at lower redshifts. 
If in addition to the
photo-z, a convolution of the distance with the photo-z error is applied to the data,
the bright-end of the LF is once more systematically enhanced (dashed lines) and can translate
into a shift of the LF towards brighter luminosities of up to one magnitude.  The
algorithms applied to the SXDS and COMBO-17 data are comparable to the latter approach,
and therefore could produce results biased towards a high space density of very bright objects.  
Notice, however, that most studies of the luminous ETG population integrate the LF over a
wide range of luminosities marked by the dark or light-grey shaded regions, and therefore
the effect on such statistics could be only minor; in particular,
the number density of galaxies brighter than $M^*$ are underestimated in the two
photo-z based $1/V_{\rm max}$ LF measurements, by only $\sim0.2$dex (see also the following Section).

The solid and dashed lines in the bottom panels, show the results from
the PML and photo-z based maximum likelihood methods (respectively).  As can be seen,
in comparison to the $1/V_{max}$ estimates from photo-zs,
the photo-z based maximum likelihood method shows a much milder tendency
of shifting the bright-end 
toward brighter magnitudes (with respect to the PML method and $1/V_{max}$ true-redshift
estimates { which provide the best and second-best matches to the true LF, respectively}).  
This brightening is almost completely counter-balanced by
a slight decrement in the space-density of faint galaxies.  
The PML method 
provides a good match to the mock result obtained using the true galaxy redshifts, and
is marginally better than the photo-z based maximum likelihood result.
Regardless of the measurement method, the recovery of the faint-end of the LF
starts to diverge from the true LF at luminosities $0.3$ to $0.6$ 
magnitudes brighter than the completeness limit
(shown by the dotted and dashed lines, respectively).  
We take this into account when studying the space density
of faint ETGs in observational catalogues.
{ In particular, for the analysis of Section 4, the space density of $z>0.6$ ETGs with $M_B(z=0)<-18.55$
could be overestimated by $50\%$.}

\subsection{Recovery of the characteristic luminosity, $M^*$}

{ Figure \ref{fig:mstarmocks} shows the recovered values of the characteristic
luminosity $M^*$ from the LFs
of the full simulation, the mocks using true redshifts, using photo-zs, and from
the double convolution with errors.  The evolution of $0.6$ magnitudes per unit
redshift is recovered in all cases and the LF from the true-redshifts is consistent
with the true underlying LF.  The brightening of the LFs affects the recovered
values of $M^*$ for the case of photo-zs and double convolution
of $\simeq 0.6$ and $1$ magnitudes at $z=1.1$, respectively.
This effect becomes
slightly less severe at lower redshifts.}

\subsection{Number density evolution with mock catalogues}

\begin{figure}
\begin{picture}(250,175)
\put(10,-100){\includegraphics[width=10.cm,clip]{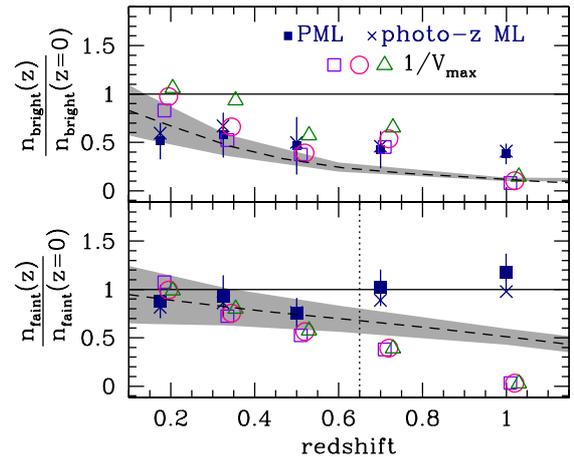}}
\end{picture}
\caption{
Ratios of ETG number densities at a given redshift
with respect to the $z\sim 0$ true luminosity function in the numerical
simulation.   The horizontal solid line shows the unit ratio.
The solid squares show the results from the mocks for the r-band from
PML estimates of the LF;
the errorbars are shown only for this case to improve clarity (and have been scaled to
an area of $0.25$sq. degrees).  { The crosses
show the results from the photo-z maximum likelihood measurement.}
The open symbols show the results from the $1/V_{max}$ weighted estimates of the mock LFs, obtained
using the true redshifts (open squares), the photometric redshifts (open circles), and the 
photo-zs convolved with their 
uncertainties (open triangles).  Small displacements along the x-axis have been added
to improve clarity.
The dashed lines correspond to the evolution inferred for the De Lucia et al. (2006) semi-analytic
model (coincident with the evolution enforced in our model galaxies; see the text).
The grey shaded area shows the estimated cosmic variance in a $0.25$sq. degree light-cone survey
divided in slices of $\Delta z=0.1$.
Top panel: samples of bright ETG galaxies. Lower panels: faint ETGs;  the dotted vertical line
shows the redshift limit for accurate faint ratios.
}
\label{fig:ratiomocks}
\end{figure}

We measure the evolution of the ETG number densities using the 
LFs from the mock catalogues to analyse the systematic
effects arising from the use of photo-z based LF measurements. 

Figure \ref{fig:ratiomocks} shows the ratios between the number density of mock ETG galaxies at
a given redshift with respect to the $z=0$ ETG LF from the full simulation box, mimicking the
analysis of Section 3.  
The dashed lines embedded in the shaded areas show the underlying evolution of this ratio and
the expected cosmic variance for the adopted cosmology, for 
spherical volumes equivalent to $\Delta z=0.1$ thick redshift slices  of light-cones 
covering a solid angle of $0.25$sq. degrees (appropriate for the ECDF-S).
In this calculation we use the Smith et al. (2003) approximation to the
non-linear power spectrum of density fluctuations.

As can be seen in the figure,
both the PML and photo-z maximum likelihood methods 
provide a good match to the underlying evolution, with a slight overestimation
at redshifts $z>0.7$ which is more severe for the faint population, as expected from the analysis
of the previous section.  Errorbars represent
the uncertainties in a field of $0.25$sq. degree area.
The other symbols show the $1/V_{max}$ estimates of the LF obtained from the true
galaxy redshifts (the best fit to the evolution in the full simulation box), 
photometric redshifts, and from photo-zs which have been
additionally convolved with the photo-z estimate errors.  As can be seen,
in the latter two cases, the bright ETG number density systematically higher although
still in agreement with the underlying evolution.

The figure also shows that for faint galaxies the systematic effect from redshift errors is negligible
compared to flux incompleteness effects which become important 
at these luminosities.  
Even if the amplitudes of the photo-z errors were
considerably larger for fainter objects 
this would still affect the faint ratios relatively little 
unless the faint-end of the LF was very steep (i.e. $\alpha<<-1$).
{ Even though, as was shown in Figure \ref{fig:mocks}, the flux incompleteness limit lies
within the limits of the faint population for the highest redshift bin,
there are important discrepancies with the true faint ETG number densities for $z\geq0.7$. 
Furthermore, the parametric fits to the LF overestimate the faint ETG densities, whereas
the $1/V_{max}$ method underestimates them, indicating an important difference between the 
two methods.}

\end{appendix}

\end{document}